\documentclass[aps,prl,twocolumn,superscriptaddress,longbibliography,nobalancelastpage]{revtex4-1}

\usepackage{graphicx}
\usepackage{amsmath}
\usepackage{color}
\usepackage[normalem]{ulem}
\usepackage[utf8]{inputenc}
\usepackage{hyperref}
\usepackage{float}
\usepackage{siunitx}
\usepackage{scalerel}

\begin{document}
\def \WSe2{WSe$_2$}
\def \MoSe2{MoSe$_2$}
\def \Vb{\textit{V}\textsubscript{b}}

\title{Overbias photon emission from light-emitting devices based on monolayer transition metal dichalcogenides}

\author{Shengyu Shan}
\thanks{These two authors contributed equally}
\affiliation{Photonics Laboratory, ETH Zürich, 8093 Zürich, Switzerland }
\author{Jing Huang}
\thanks{These two authors contributed equally}
\affiliation{Photonics Laboratory, ETH Zürich, 8093 Zürich, Switzerland }
\author{Sotirios Papadopoulos}
\affiliation{Photonics Laboratory, ETH Zürich, 8093 Zürich, Switzerland }
\author{Ronja Khelifa}
\affiliation{Photonics Laboratory, ETH Zürich, 8093 Zürich, Switzerland }
\author{Takashi Taniguchi}
\affiliation{International Center for Materials Nanoarchitectonics, National Institute for Materials Science, 1-1 Namiki, Tsukuba 305-0044, Japan}
\author{Kenji Watanabe}
\affiliation{Research Center for Functional Materials, National Institute for Materials Science, 1-1 Namiki, Tsukuba 305-0044, Japan}
\author{Lujun Wang}
\affiliation{Photonics Laboratory, ETH Zürich, 8093 Zürich, Switzerland }
\author{Lukas Novotny}
\email{lnovotny@ethz.ch}
\affiliation{Photonics Laboratory, ETH Zürich, 8093 Zürich, Switzerland }
 
\date{\today}

\begin{abstract}
Tunneling light-emitting devices (LEDs) based on transition metal dichalcogenides (TMDs) and other 2D materials are a new platform for on-chip optoelectronic integration. Some of the physical processes underlying this LED architecture are not fully understood, especially the emission at photon energies higher than the applied electrostatic potential, so-called overbias emission.
Here we report overbias emission for potentials that are near half of the optical bandgap energy in TMD-based tunneling LEDs. We show that this emission is not thermal in nature, but consistent with exciton generation via a two-electron coherent tunneling process. 

\end{abstract}

\maketitle
\section{\label{sec:Intro}Introduction}

\par In 2015, the first 2D material-based tunneling light-emitting device (LED) was realized~\cite{Withers2015,Withers2015b}. It employed graphene (Gr) as a conductor for electrical contacts, transitional metal dichalcogenides (TMDs) as semiconductors, and hexagonal boron nitride (hBN) as an insulator. This LED architecture has inspired investigations on cavity integration~\cite{Liu2017NanocavityDiode, DelPozo-Zamudio2020ElectricallyMicrocavities}, single defect LEDs~\cite{Clark2016SingleHeterostructure} and exciton modulation~\cite{Ryu2021}. It also opened up a new perspective for integrated on-chip optoelectronic devices~\cite{Mak2016}.

\par A typical device architecture is shown in Fig.~\ref{fig:Origin}a. It consists of a Gr-hBN-WSe\textsubscript{2}-hBN-Gr heterostructure, with two monolayer Gr flakes acting as transparent electrodes and two hBN multilayers defining the tunnel barriers. A monolayer \WSe2 is sandwiched in the middle and serves as the active material. Such double-tunnel barrier LEDs provide large-area exciton light emission with an external quantum efficiency (EQE) on the order of $10^{-2}$ at room temperature \cite{Withers2015,Withers2015b}. Here, excitons are formed by charge injection of both electrons and holes into the active layer. This requires the applied bias potential ($eV_\mathrm{b}$, where $e$ is the elementary charge, $V_\mathrm{b}$ is the bias voltage) to be larger than the optical bandgap energy so that electrons and holes can tunnel from the Gr electrodes to \WSe2, thereby forming excitons~\cite{Binder2017Sub-bandgapHeterostructure}. 
\par However, there are also alternative ways to generate excitons for  light emission, such as by energy transfer. This process involves inelastic electron tunneling (IET), in which the electron couples its energy to TMD excitons during the tunneling process \cite{Papadopoulos2022,Pommier2019ScanningSemiconductor,PenaRoman2022ElectroluminescenceLight}. Such energy transfer can occur efficiently in van der Waals (vdW) heterostructures and is due to strong near-field coupling between the tunneling electrons and the active material. Thus, excitons in TMDs can be generated either by charge injection or by energy transfer. In both processes energy conservation requires that the bias potential $eV_\mathrm{b}$ is larger than the optical bandgap energy $\hbar\omega_{\mathrm{BG}}$ ($\hbar\omega_{\mathrm{BG}}$ $\simeq$ \SI{1.64}{eV} for monolayer \WSe2 at room temperature~\cite{Kozawa2014PhotocarrierDichalcogenides}, where $\hbar$ is the reduced Planck constant, and $\omega_\mathrm{BG}$ is the angular transition frequency); no excitonic photon emission is expected for $eV_\mathrm{b}<\hbar\omega_{\mathrm{BG}}$ \cite{Binder2017Sub-bandgapHeterostructure,Papadopoulos2022}.

\par In {\em optically} excited systems, photon emission at energies larger than the excitation energy can be generated by second-order processes, such as two-photon excitation~\cite{He2014TightlyWSe2}. In {\em electrically} pumped systems, such upconversion can be facilitated by an intermediate state, for example by Auger scattering of interlayer excitons~\cite{Binder2019UpconvertedHeterostructures}. In light-emitting junctions apart from vdW heterostructures, overbias emission can also be generated by thermally assisted upconversion~\cite{Pechou1998CutoffAir,Su2022ThermalDiodes}, non-thermal-equilibrium carrier generation~\cite{Buret2015SpontaneousAntennas,Zhu2020Hot-carrierJunctions,Cui2020ElectricallyJunctions,Lian2022Ultralow-voltageDiodes} and coherent multielectron processes~\cite{Xu2014OverbiasJunction,Xu2016DynamicalJunction,Peters2017,Fung2020TooProcesses,Zhu2022TuningJunctions}.

\par In this paper, we report on exciton light emission from a monolayer TMD tunneling LED driven by bias potentials ($eV_\mathrm{b}\simeq$ \SI{1.00}{\electronvolt}) much smaller than the optical bandgap energy ($\hbar\omega_{\mathrm{BG}}\simeq$ \SI{1.64}{\electronvolt}). To identify the physical origins of this overbias emission we perform electroluminescence (EL) measurements on various LED designs and at different temperatures.

\par In addition to double-barrier LEDs we also investigate single-barrier Gr-TMD-hBN-gold  heterostructures, with TMD = \{\WSe2, \MoSe2\}. Compared with double-barrier LEDs, single-barrier LEDs can reach higher currents under the same bias voltage, thus allowing us to observe exciton emission at very low bias voltages. With this architecture, we start to detect light emission from the A-exciton in \WSe2 at \SI{0.81}{V} and at \SI{0.74}{V} in \MoSe2. The measured threshold voltages correspond to approximately half the optical bandgap energies. This observation hints at a second-order energy transfer process  based on multielectron tunneling~\cite{Xu2014OverbiasJunction,Xu2016DynamicalJunction,Peters2017}.

\section{\label{sec:DB LED} Overbias Light emission FROM A double-barrier LED}

\begin{figure}[!hbt]
\centering
\includegraphics[width=1\columnwidth]{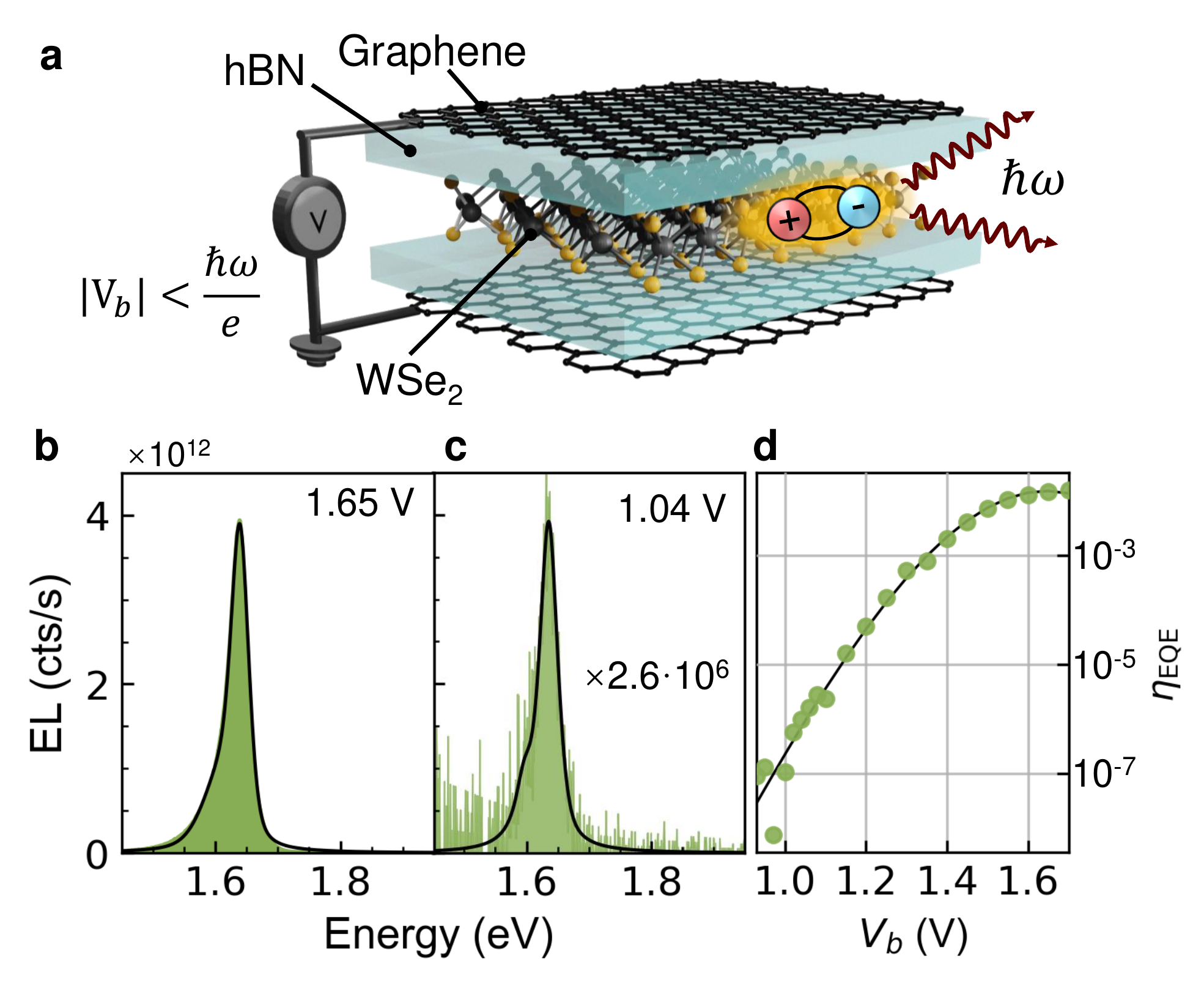}
\caption{\textbf{a}, Illustration of a double-barrier tunneling LED. The junction is encapsulated in hBN on both sides (not shown). \textbf{b}, \textbf{c}, EL spectra of the double-barrier LED for V\textsubscript{b} = \SI{1.65}{\volt} and V\textsubscript{b} = \SI{1.04}{\volt}, respectively. The measured spectra (green areas) are fitted with the sum of two pseudo-Voigt functions (black lines) representing A-exciton and trion. \textbf{d}, EQE (in the spectral range from 1.4 to \SI{1.8}{e\volt}) as a function of applied bias. The green dots represent data points and the black curve is a guide-to-the-eye.} \label{fig:Origin}
\end{figure}
\par We first describe our results for the double-barrier LED shown in Fig.~\ref{fig:Origin}a.  The core structure is a vertical assembly of Gr-hBN-\WSe2-hBN-Gr, in which two Gr flakes serve as electrodes. The hBN thickness corresponds to 4$\pm$1 atomic layers. This tunnel junction is encapsulated between two thick hBN flakes. We fabricate our devices by using the dry pick-up and transfer method~\cite{Zomer2014}, where we transfer the entire device onto a glass coverslip. After transfer we fabricate edge contacts to the two graphene electrodes ~\cite{Wang2013,Overweg2018}. EL is collected with an oil-immersion objective from the glass side and detected by a spectrometer. (See the Supplemental Material \cite{[{See Supplemental Material at }][{ which includes Refs.~[2, 9, 17, 19, 20, 23, 25-28, 31-33, 37, 38, 42].}]SI}, Sec.~I, II).


\par Monolayer \WSe2 has an electronic bandgap of $\sim$\SI{1.82}{e\volt}~\cite{Gutierrez-Lezama2021IonicSemiconductors} and an optical bandgap of $\sim$\SI{1.64}{e\volt} at room temperature~\cite{Kozawa2014PhotocarrierDichalcogenides}. To electrically generate excitons, the bias potential $eV_\mathrm{b}$ has to be larger than the optical bandgap energy~\cite{Binder2017Sub-bandgapHeterostructure}. Figure~\ref{fig:Origin}b shows a representative EL spectrum for $V_\mathrm{b}$ = \SI{1.65}{\volt}. The peak of the spectrum centers at $\sim$\SI{1.64}{e\volt}, which corresponds to the A-exciton of \WSe2~\cite{Binder2017Sub-bandgapHeterostructure}. The asymmetric broadening at lower energy can be associated with the trion~\cite{Binder2017Sub-bandgapHeterostructure}. However, we also observe exciton light emission for $eV_\mathrm{b}$ significantly smaller than the optical bandgap. As an example, Fig.~\ref{fig:Origin}c shows the EL spectrum for $V_\mathrm{b}$ = \SI{1.04}{\volt}. Compared to Fig.~\ref{fig:Origin}b, this spectrum has the same main peak position and similar linewidth, indicating that the spectrum is also dominated by the contribution from A-exciton. The spectral shape remains similar, but the intensity and hence the EQE decreases. We define EQE as $\eta_{\text{EQE}}=\frac{\Gamma^\mathrm{X}}{I/e}$, where $\Gamma^\mathrm{X}$ is the photon count rate in the spectral range from 1.4 to \SI{1.8}{e\volt} and $I$ is the electrical current (for more details, see the Supplemental Material {\cite{SI}}, Sec.~III). As shown in Fig.~\ref{fig:Origin}d, the EQE drops exponentially with decreasing $V_\mathrm{b}$ and disappears in the noise floor at $\sim$\SI{0.93}{\volt}. To extend the measurement range to even lower bias voltages we require a higher emission intensity and hence a higher tunnel current. Therefore, in a next step, we eliminate one of the tunnel barriers and repeat the measurements for a single-barrier device.

\section{\label{sec:SB LED}Overbias light emission from a \WSe2-based single-barrier LED}
\begin{figure}[!hbt] 
\centering
\includegraphics[width=1\columnwidth]{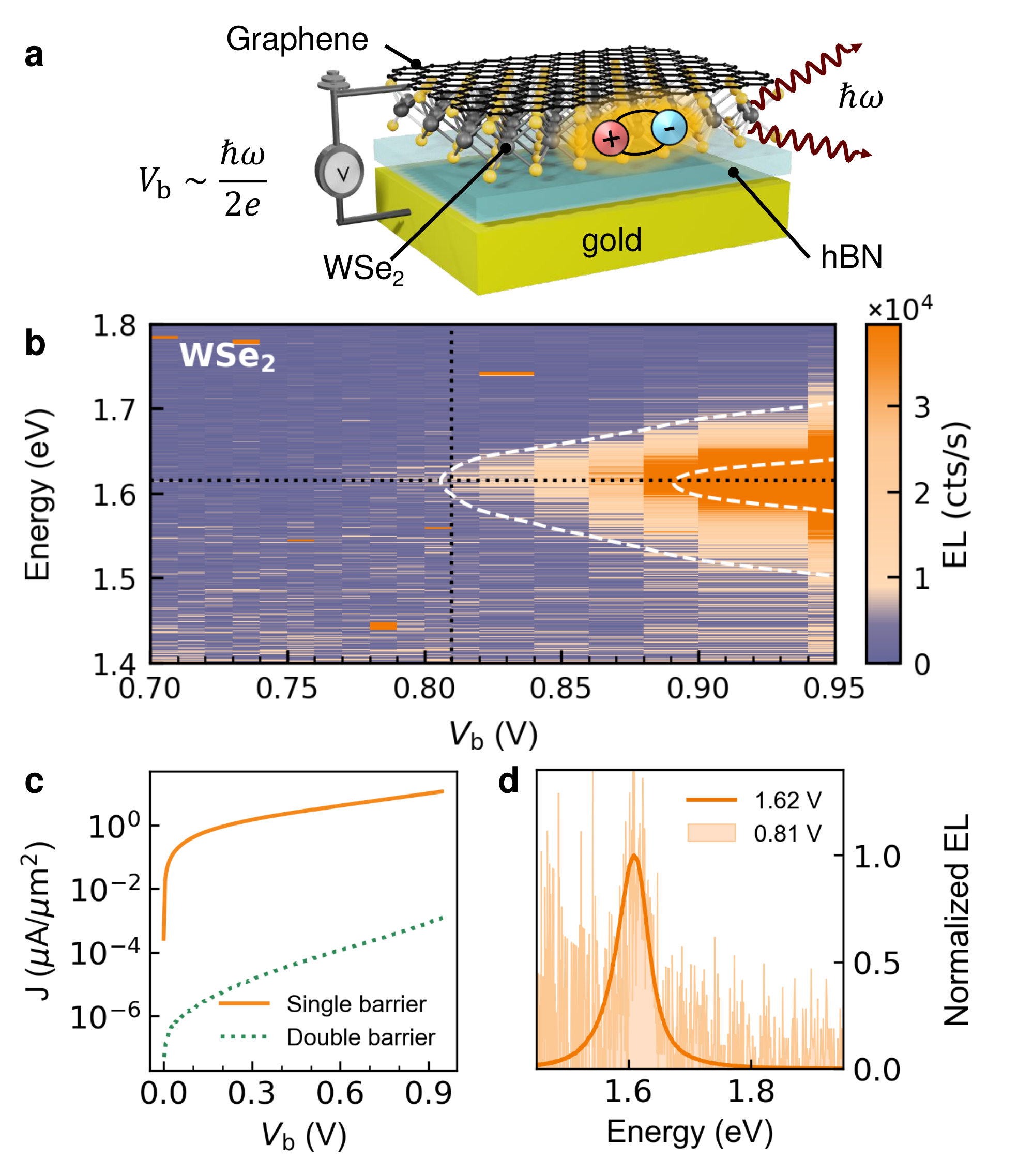}
\caption{\textbf{a}, Illustration of a single-barrier LED. The stack is encapsulated by a top hBN flake (not shown). \textbf{b}, EL spectra for $V_\mathrm{b}$ ranging from \SI{0.7}{\volt} to \SI{0.95}{\volt}. The horizontal dotted line indicates the neutral exciton energy (1.62 eV) and the vertical dotted line denotes the threshold for exciton emission. There is a factor of 10 difference between the two dashed contour lines. \textbf{c}, Current density-voltage (J-V) curve of a single- and a double-barrier LED. \textbf{d}, Normalized EL spectra for $V_\mathrm{b}$ = \SI{0.81}{\volt} and \SI{1.62}{\volt}.}  \label{fig:SinBarWSe2}
\end{figure} 
The architecture of a single-barrier LED is shown in Fig.~\ref{fig:SinBarWSe2}a. The device is composed of a Gr-\WSe2-hBN-gold heterostructure, where the monolayer Gr is contacted to a gold electrode. As shown in Fig.~\ref{fig:SinBarWSe2}c, by using a single-barrier device (hBN with 3$\pm$1 atomic layers) we are able to increase the current density by $\sim$4 orders of magnitude over the previous double-barrier device. The EL spectra of the single-barrier device are shown in Fig.~\ref{fig:SinBarWSe2}b for different bias voltages. The spectra have a peak at $\sim$\SI{1.62}{eV} (horizontal dotted line), which is slightly red-shifted compared to the double-barrier LED. Consequentially, we assign this peak to the neutral A exciton, which is shifted to lower energies due to the stronger dielectric screening of the directly contacting Gr~\cite{Lorchat2020FilteringGraphene}. The overall EL is moderately quenched and becomes trion-free due to both charge and energy transfer~\cite{Lorchat2020FilteringGraphene,Froehlicher2018ChargeHeterostructures}. As we gradually lower $V_\mathrm{b}$, the exciton peak remains visible in the spectrum, even for $eV_\mathrm{b}$ = \SI{0.81}{\electronvolt} (vertical dotted line), corresponding to half of the \WSe2 optical bandgap energy ($\hbar\omega_{\mathrm{BG}}$ = \SI{1.62}{e\volt}). The EL spectrum for $V_\mathrm{b}=\SI{0.81}{\volt}$ is shown in Fig.~\ref{fig:SinBarWSe2}d (light orange area). Its shape is almost identical to the spectrum recorded for \SI{1.62}{\volt} (solid orange curve). This observation hints at a second-order process involving two electrons.

\section{\label{sec:MoSe2}Overbias light emission from a \MoSe2-based single-barrier LED}

\begin{figure}[!tb] 
\centering
\includegraphics[width=1\columnwidth]{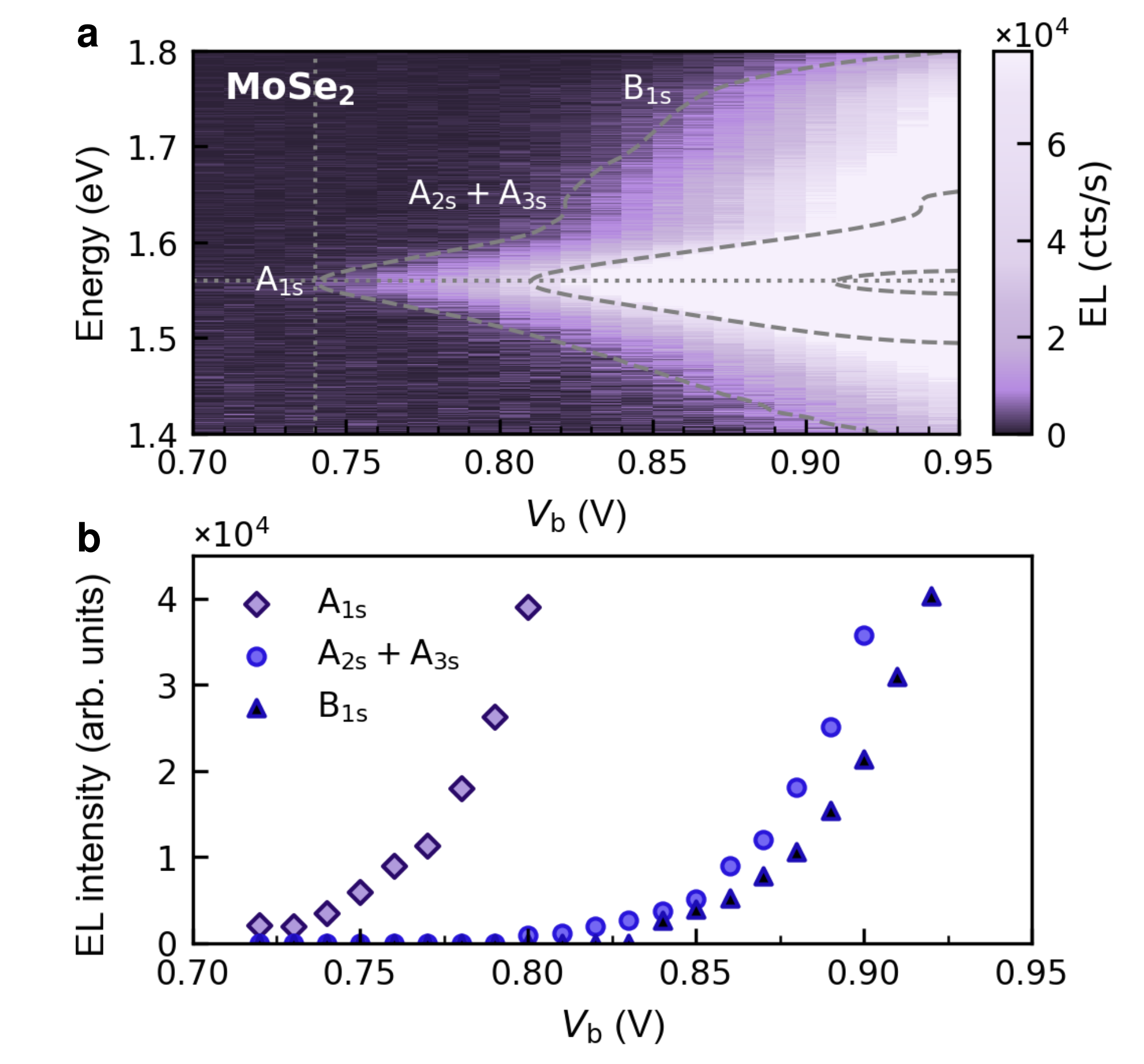}
\caption{\textbf{a}, EL spectra of a single-barrier LED based on \MoSe2. The horizontal dotted line indicates the neutral exciton energy (\SI{1.56}{\electronvolt}) and the vertical dotted line denotes the threshold for exciton emission. There is a factor of 10 between adjacent dashed contour lines. \textbf{b}, Dependence of the integrated EL intensity of $\mathrm{A}_{\mathrm{1s}}$, $\mathrm{A}_{\mathrm{2s}}+\mathrm{A}_{\mathrm{3s}}$, and $\mathrm{B}_{\mathrm{1s}}$ excitons on bias voltage.}  \label{fig:SinBarMoSe2}
\end{figure}

In order to further strengthen our interpretation, we replace \WSe2 by \MoSe2, which has a lower bandgap and should therefore lead to EL at even lower bias voltages. Furthermore, it is known that the exciton emission from \MoSe2 is less affected by Gr quenching~\cite{Lorchat2020FilteringGraphene}, thus yielding stronger EL emission and providing a better signal-to-noise ratio. Figure \ref{fig:SinBarMoSe2}a shows voltage-dependent EL spectra, in which the peak near \SI{1.56}{eV} (horizontal dotted line) is assigned to the red-shifted A-exciton (1s state) of monolayer \MoSe2~\cite{Lorchat2020FilteringGraphene}. This feature appears at the lowest voltage of \SI{0.74}{\volt} (vertical dotted line), which again is much lower than the photon energy of \SI{1.56}{V}. At higher biases, two side peaks appear near \SI{1.66}{eV} and \SI{1.71}{eV}. According to their energy offsets relative to the A-exciton we assign the first to the 2s and 3s state of A-exciton and the second to the B-exciton~\cite{Han2018}. 

To analyze the voltage dependence of these three features, we fit the spectra with three pseudo-Voigt functions. The corresponding fitting amplitudes are plotted in Fig.~\ref{fig:SinBarMoSe2}b as a function of the bias voltage (see the Supplemental Material {\cite{SI}}, Sec.~IV). We observe that the three peaks emerge at different bias voltages: the lowest state of A-exciton with a peak position near \SI{1.56}{eV} appears for $V_\mathrm{b}>$~\SI{0.74}{\volt}, the 2s and 3s excited states near \SI{1.66}{eV} have an onset voltage of \SI{0.82}{\volt}, and the B-exciton with the highest energy ($\sim$\SI{1.71}{eV}) emerges at $V_b=$~\SI{0.84}{\volt}. Altogether, each of the three features in \MoSe2 emerges at bias potentials of half the photon energy (e$V_\mathrm{b}\simeq\hbar\omega/2$), similar to the \WSe2 device.

\section{\label{sec:mechanism}Analysis of underlying mechanisms}
Besides a second-order process involving two electrons, other processes can also give rise to overbias emission. These include:
\begin{enumerate}
    \item Blackbody radiation of hot carriers, in which the effective temperature is related to the bias voltage or the input power~\cite{Buret2015SpontaneousAntennas,Joulain2003,Greffet2018LightLaw,Zhu2020Hot-carrierJunctions,Cui2020ElectricallyJunctions}.
    \item Recombination of out-of-equilibrium carriers~\cite{Lian2022Ultralow-voltageDiodes}, in which electrons and holes in the high energy tail of the Fermi-Dirac distribution tunnel into the TMD to form excitons.
    \item Second-order nonlinear optical processes, in which photons generated by IET \cite{Kuzmina2021,Parzefall2019} excite excitons in the TMD.
    \item Second-order energy transfer, in which the energy from pairs of coherently tunneling electrons~\cite{Xu2014OverbiasJunction,Xu2016DynamicalJunction,Peters2017} is forming excitons in the TMD.
\end{enumerate}

To exclude the first two effects, we fabricated yet another single-barrier \MoSe2 LED and measured its EL at cryogenic temperature ($\sim$\SI{10}{\kelvin}). (See the Supplemental Material {\cite{SI}}, Sec.~V). To rule out a thermal origin for the observed overbias emission we use the following blackbody radiation model for the radiated power~\cite{Buret2015SpontaneousAntennas,Cui2020ElectricallyJunctions,Zhu2020Hot-carrierJunctions}:
\begin{equation}\label{eq:Pther}
    P_\mathrm{ther} =\int_{0}^{\infty}\frac{\omega^2}{\pi^2 c^3}\frac{\hbar \omega}{\mathrm{exp}({\hbar\omega/k_\mathrm{B} T^{\prime}})-1}\epsilon^{\prime\prime}(\omega)d\omega,
\end{equation}
\noindent where $c$ is the speed of light, $\omega$ the photon angular frequency, $k_\mathrm{B}$ the Boltzmann constant, $T^{\prime}$ the effective hot carrier temperature and $\epsilon^{\prime\prime}$ the emissivity of the TMD exciton, which can be derived from the refractive index~\cite{Liu2020Temperature-dependentCalculations}. For resistive heating we obtain the linear dependence~\cite{Cui2020ElectricallyJunctions,Zhu2020Hot-carrierJunctions}:
\begin{equation}\label{eq:Tdevice}
T^{\prime} = T_0 + \kappa \frac{e}{k_\mathrm{B}} V_\mathrm{b},
\end{equation}
where $T_0$ is the lattice temperature and $\kappa$ is a temperature-independent dimensionless constant that can be derived from experimental data at room temperature. With this $\kappa$, Eq.~(\ref{eq:Pther}) predicts that the radiated power in the spectral region of the exciton should decrease by roughly 9 orders of magnitude when $T_0$ is reduced from \SI{300}{K} to \SI{10}{K}. However, our measurements show only a decrease of less than 2 orders of magnitude. This huge discrepancy between model and measurement indicates that blackbody radiation is not the source of the observed overbias emission. The same is true for the second scenario, the recombination of out-of-equilibrium carriers, since our measurements reveal that the dependence of the radiated power on bias voltage is unaffected by the lattice temperature. (See the Supplemental Material {\cite{SI}}, Sec.~V for analysis details).

The third scenario involves two steps, namely photon emission by IET~\cite{Parzefall2019,Kuzmina2021} and a subsequent nonlinear optical process. Comparing the photon emission efficiencies of IET and the observed overbias emission, we require a nonlinear optical process with unit efficiency to explain our measurements. Therefore, it is safe to discard the third scenario as an explanation for our observation.

We are left with the fourth scenario, illustrated in Fig.~\ref{fig:2e_process}a. In this scenario, excitons are generated by the action of two electrons. This process is supported by two recent observations. First, it has been demonstrated that excitons can be efficiently excited by tunneling electrons via nonradiative energy transfer~\cite{Papadopoulos2022}. Second, it has been reported that multielectron coherent tunneling can generate overbias emission in plasmonic tunnel junctions~\cite{Xu2014OverbiasJunction,Xu2016DynamicalJunction, Peters2017,Fung2020TooProcesses,Zhu2022TuningJunctions}. Therefore, we identify multielectron IET as the most likely mechanism responsible for the observed overbias emission.

\section{\label{sec:citeref}Theory of two-electron energy transfer}
\begin{figure}[!t]
\includegraphics[width=1\columnwidth]{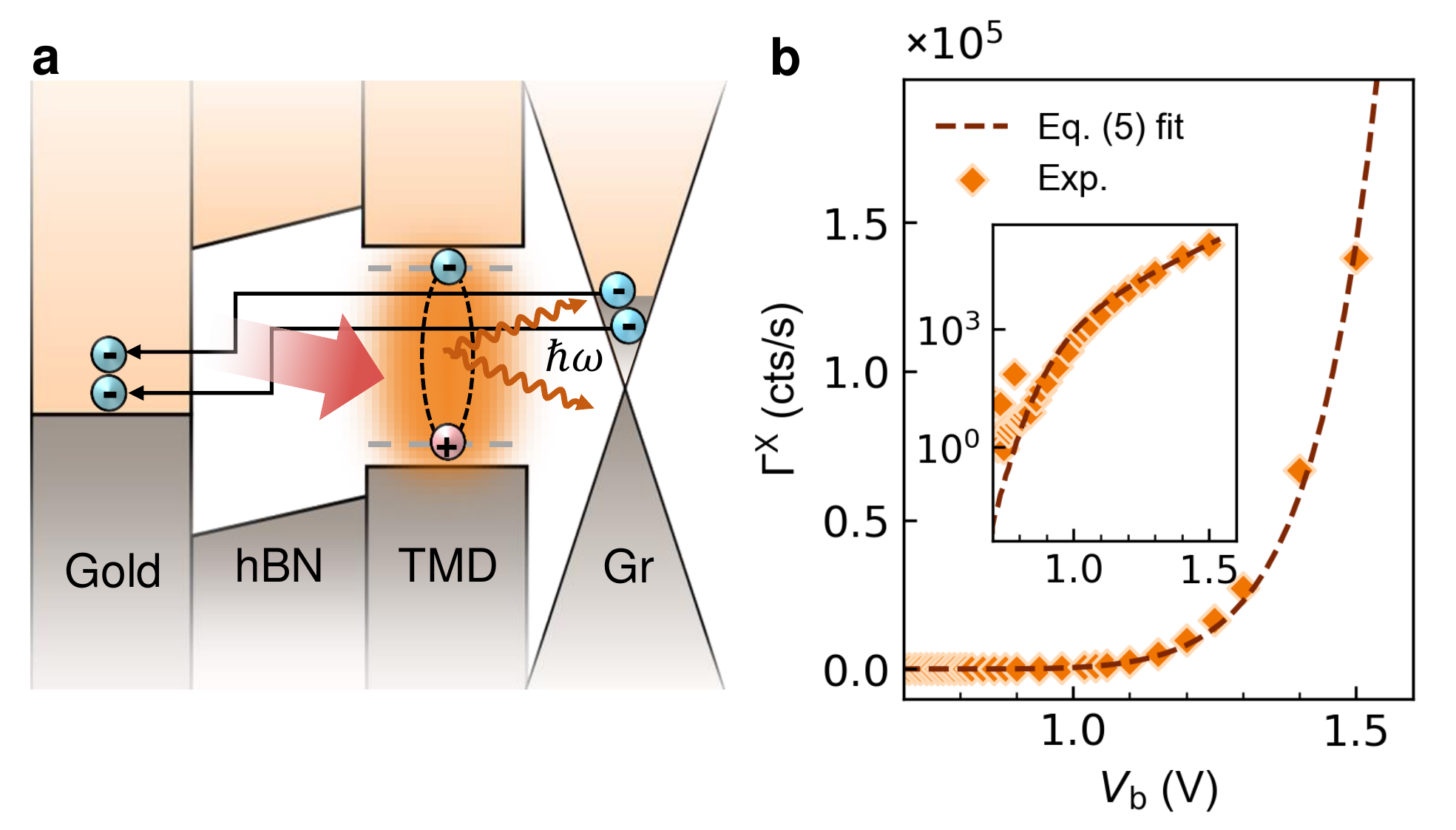}
\caption{\textbf{a}, Energy transfer based on two-electron coherent tunneling. A pair of electrons tunnel inelastically and their combined energy generates excitons in the TMD. \textbf{b}, Exciton EL intensity ($\Gamma^\mathrm{X}$) of a single-barrier \WSe2 LED as a function of bias voltage $V_\mathrm{b}$. The inset shows the data on a semi-logarithmic scale. The data points correspond to the integrated EL photon count rate in the spectral range from 1.4 to \SI{1.8}{eV}. The dashed curve is the fitting result of Eq.~(\ref{eq:Ix}).}  \label{fig:2e_process}
\end{figure}\par

In plasmonic tunnel junctions, overbias light emission based on two-electron IET depends on the interplay between higher-order quantum noise and the local density of optical states (LDOS)~\cite{Xu2014OverbiasJunction,Xu2016DynamicalJunction}. Here we adopt this theory to a TMD-coupled tunnel junction. The non-symmetrized power spectral density of the fluctuating tunnel current reads as~\cite{Roussel2016PerturbativeCircuits,Fevrier2018TunnelingNoise}
\begin{equation}\label{eq:Si}
\begin{split}
S_{ii}(\omega,V_\mathrm{b}) = &e\{[1+n_{B}(eV_\mathrm{b}-\hbar\omega)]I(V_\mathrm{b}-\hbar\omega/e) \\ 
& +n_{B}(eV_\mathrm{b}+\hbar\omega)I(V_\mathrm{b}+\hbar\omega/e) \},
\end{split}
\end{equation}
\noindent where $I(V_\mathrm{b})$ is the bias-dependent tunnel current and $n_{B}(x) = (\mathrm{exp}(x/k_\mathrm{B}T)-1)^{-1}$ is the Bose-Einstein distribution at temperature T. We are concerned with the absorption of electromagnetic energy generated by the fluctuating tunneling current, which is described by the positive frequency part of $S_{ii}$ \cite{blanter2000shot}.

The absorption depends on the local environment of the tunnel junction, and is mathematically described by the LDOS ($\rho$) and the system's Green's function~\cite{Novotny2009Nanooptics}. For frequencies near the TMD exciton the absorption is dominated by the LDOS of the TMD ($\rho_\mathrm{TMD}$). In a two-electron process, the locally absorbed energy is no longer linearly dependent on $S_{ii}$. In analogy to previous studies \cite{Peters2017,Fung2020TooProcesses,Zhu2022TuningJunctions} the two-electron absorption rate $\gamma_{2e}$ can be represented as
\begin{equation}\label{eq:P2e}
\begin{split}
    \gamma_{2e}(\omega,V_\mathrm{b}) \propto \rho_\mathrm{TMD}(\omega) \int_{0}^{\omega} & \rho_\mathrm{TMD}(\omega')S_{ii}(\omega',V_\mathrm{b}) \\ 
& S_{ii}(\omega-\omega',V_\mathrm{b})d\omega',
\end{split}
\end{equation}
\noindent where $\rho_\mathrm{TMD}$ is calculated by following Ref.~\cite{Parzefall2019}. Equation~(\ref{eq:P2e}) describes a two-electron tunneling process, in which the energy of two electrons is absorbed by the TMD to generate an exciton (Fig.~\ref{fig:2e_process}a). Since excitons can only be generated by energies larger than the exciton energy ($\hbar\omega>E_\mathrm{X}$) we can represent the exciton emission intensity $\Gamma^\mathrm{X}$ as

\begin{equation}\label{eq:Ix}
    \Gamma^\mathrm{X}(V_\mathrm{b}) \propto \int_{E_\mathrm{X}/\hbar}^{\infty} \gamma_{2e}(\omega,V_\mathrm{b}) d\omega.
\end{equation}

\noindent As can be seen in Fig.~\ref{fig:2e_process}b, the exciton EL intensity increases exponentially with increasing $V_\mathrm{b}$, and the calculated $\Gamma^\mathrm{X}(V_\mathrm{b})$ according to Eq.~(\ref{eq:Ix}) agrees well with the experimental results. (See the Supplemental Material {\cite{SI}}, Sec.~VI). This agreement supports our interpretation that the overbias emission in our TMD-based LEDs results from  two-electron tunneling followed by energy transfer. 

In summary, we investigated exciton light emission for potentials lower than the optical bandgap energy in TMD-based tunneling LEDs. We are able to measure exciton emission for bias  potentials of only half the optical bandgap energy. We explain this overbias emission by a second-order energy transfer process.

{\em Acknowledgments---}
This work has been supported by the Swiss National Science Foundation (grant 200020\_192362/1). The authors are grateful to Olivier~Huber, Deepankur~Thureja, Atac~Imamoglu and Jian~Zhang for kindly helping us perform the cryogenic measurements. We acknowledge Hsiang-Lin Liu for providing us with the original data of TMD optical constants from Ref.~\cite{Liu2020Temperature-dependentCalculations}. We also thank Antti~Moilanen, Anna~Kuzmina, Achint~Jain, Yesim~Koyaz, Yang~Xu, Nicola~Carlon~Zambon, Moritz~Cavigelli, Martin Frimmer, Jonas David Ziegler and Giacomo~Scalari for fruitful discussions and support. 
The use of the facilities of the FIRST center for micro- and nanoscience at ETH Zürich is gratefully acknowledged. 
K.W. and T.T. acknowledge support from JSPS KAKENHI (Grant Numbers 19H05790, 20H00354 and 21H05233).

L.N., L.W. and S.P. conceived the project. J.H., S.S. and R.K. fabricated the devices and performed the experiments. L.N., L.W. and S.P. supervised the project. T.T. and K.W. synthesized the h-BN crystals. S.S, J.H. and L.N. analyzed the data and co-wrote the manuscript.

\bibliographystyle{apsrev4-1_abbrv}
\bibliographystyle{apsrev4-1_abbrv}
\bibliography{References}

\end{document}